\PassOptionsToPackage{dvipsnames}{xcolor}

\documentclass[%
 reprint,
 superscriptaddress,
 amsmath,amssymb,
 aps
]{revtex4-2}

\usepackage{graphicx, nicefrac}
\usepackage{xcolor}
\usepackage{dcolumn}
\usepackage{bm}
\usepackage{wasysym}

\newcommand{\mUP}{{\color{cyan}$\blacktriangle$}}
\newcommand{\mDOWN}{{\color{LimeGreen}$\blacktriangledown$}}
\newcommand{\mZERO}{{\color{gray}\CIRCLE}}

\begin{document}

\preprint{APS/123-QED}

\title{Critical dynamics and phase transition of a strongly interacting warm spin-gas}

\author{Yahel Horowicz}
\thanks{These authors contributed equally to this work.}
\affiliation{Department of Physics of Complex Systems, Weizmann Institute of Science, Rehovot 76100, Israel}

\author{Or Katz}
\thanks{These authors contributed equally to this work.}
\affiliation{Department of Physics of Complex Systems, Weizmann Institute of Science,
Rehovot 76100, Israel}
\address{present address: Department of Electrical and Computer Engineering, Duke University, Durham, North Carolina 27708, USA}
\email[Corresponding author: ]{or.katz@duke.edu}

\author{Oren Raz}
\affiliation{Department of Physics of Complex Systems, Weizmann Institute of Science, Rehovot 76100, Israel}

\author{Ofer Firstenberg}
\affiliation{Department of Physics of Complex Systems, Weizmann Institute of Science, Rehovot 76100, Israel}

\date{\today}

\begin{abstract}
Phase transitions are emergent phenomena where microscopic interactions drive a disordered system into a collectively ordered phase. Near the boundary between two phases, the system can exhibit critical, scale-invariant behavior. Here, we report on a second-order phase transition accompanied by critical behavior in a system of warm cesium spins driven by linearly-polarized light. The ordered phase exhibits macroscopic magnetization when the interactions between the spins become dominant. We measure the phase diagram of the system and observe the collective behavior near the phase boundaries, including power-law dependence of the magnetization and divergence of the susceptibility. Out of equilibrium, we observe a critical slow-down of the spin response time by two orders of magnitude, exceeding five seconds near the phase boundary. This work establishes a controlled platform for investigating equilibrium and nonequilibrium properties of magnetic phases.
\end{abstract}

\maketitle


\section{Introduction}
Investigation of correlated states of matter and their macroscopic phases are at the frontier of inter-disciplinary physical research. The macroscopic phase of interacting particles is determined by an interplay between the energy and the entropy of the system. When the entropy dominates, the system becomes disordered whereas, for energy dominated system, the ordered phase dominates as it minimizes the free energy \cite{Mermin502, StatisticalMechanics}. Various phases are commonly described by macroscopic order-parameters such as density, conductivity, or magnetization. The transition between different phases is commonly accompanied by non-analytic behavior of some properties of the system, \textit{e.g.}, the susceptibility, correlation length, and time. These non-analytic properties are often associated with critical exponents, which can be categorized into universality classes that depend only on robust properties, such as the dimensionality and symmetries of the system \cite{Mermin502, StatisticalMechanics,PT_Schneider_1973}. Although first discussed in the context of equilibrium statistical mechanics, phase transitions with characteristic critical exponents  commonly appear also in driven, non-equilibrium systems, and the properties of these transitions are often quite different from those of equilibrium systems \cite{marro2005nonequilibrium,henkel2008non}.  

Spin systems are used as key examples for magnetic phase transitions as they are often relatively easy to study \cite{Glauber1963time, blundell2001magnetism, Carrasquilla2017, Gyorffy_1985, collins1989magnetic}. In condensed matter systems, strong interactions between neighboring spins can overcome their entropy and consequent with ordered magnetic phases, as manifested in the rich phase diagrams of a wide range of materials and temperatures \cite{FERR, Sato49, ATIQ20155262,pt_Iron}. In gaseous systems in contrast, the particles typically interact sporadically, leading to prevalence of entropy over the interaction energy and consequent with a magnetically disordered phase. Only at ultra-cold temperatures in which entropy is sufficiently low and quantum effects become dominant, atomic gases can exhibit magnetically ordered phase \cite{Jo1521, Simon2011}. As atomic gases feature both long spin lifetimes and high degree of control by optical means, they have prominent applications in physical studies of equilibrium and non-equilibrium critical phenomena \cite{Mazurenko2017, PhysRevX.8.021069, Bernien2017, Gross995}. However, non-trivial phases of optically controlled spin gases at ambient conditions have never been characterized.

The specific system we study in this manuscript is a vapor of neutral alkali atoms, above room temperature. These atoms have a nonzero spin at their electronic ground level and could thus sustain steady magnetization. The spin, a composite of the electronic and nuclear spins, can be prepared, controlled, and monitored by optical means utilizing the strong spin-orbit coupling provided by the single valence electron \cite{happer2010optically,auzinsh2010optically}. Frequent spin-exchange collisions between pairs of atoms in the vapor manifest a local spin-dependent interaction. This interaction often leads to decoherence and relaxation \cite{Novikova2012} but can also enable coherent coupling and facilitate optical pumping and sensing in various applications \cite{chalupczak2012enhancement, romalis2010hybrid, serf1977, katz2019quantum, kominis2003subfemtotesla, HybridHe3, Katz2018, mouloudakis2020spin, kong2020measurement, CoherentCoupling2015, Pfau446,WANG201827, LongEntanglement2020}. In particular, a pioneering work by Forston \textit{et al.} has demonstrated the emergence of spontaneous spin-polarization and magnetic bi-stability upon absorption of linearly-polarized light \cite{SP87, SP96,SP2002}. Forston \textit{et al.} have characterized the spin state by measuring the hysteresis appearing for slow variation of the light polarization. However, the critical and collective behavior at the conditions where spontaneous polarization occurs has not been systematically studied.

\begin{figure}[t!]
\begin{centering}
 	\includegraphics[width=8.7 cm]{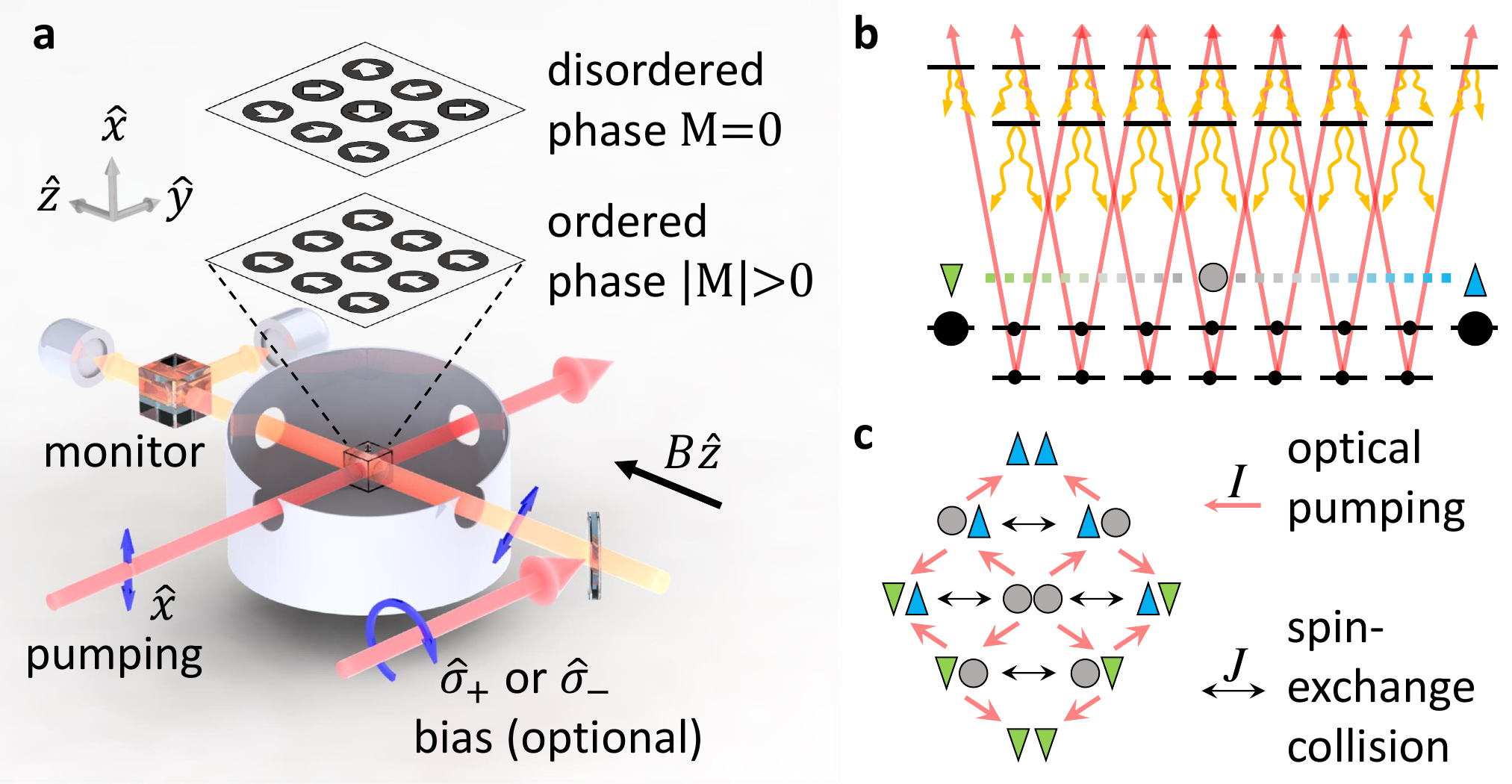} 
\end{centering}\centering{}\setlength{\belowcaptionskip}{-4mm}\caption{\textbf{Experimental system and coupling scheme.} (a) Cesium atoms contained in a glass cell experience frequent local spin-exchange collisions and are illuminated by linearly-polarized pumping light. The pumping, with rate $I$, aligns each spin symmetrically either parallel or anti-parallel to the magnetic field $B\hat{z}$. The collisions, with rate $J$, enable the generation of local correlations that lead to the ordering of the spins at high densities.
A weak, far detuned, monitor beam is used to measure the total atomic magnetization $M$ along $\hat{z}$, which acts as an order parameter. For measurements of the magnetic susceptibility, an auxiliary, circularly-polarized, light beam is introduced to bias the optical pumping toward positive $M$ ($\hat{\sigma}_+$ bias beam) or negative $M$ ($\hat{\sigma}_-$ bias beam). (b) Atomic level structure. In the $\hat{z}$ quantization basis, the pumping light comprises $\hat{\sigma}_+$ and $\hat{\sigma}_-$ components (red arrows). It does not excite the maximally polarized states (\mUP~and \mDOWN, corresponding to $M=\pm1$). In conjunction with spontaneous emission (orange arrows), the pumping drives the unpumped atoms ($|M|<1$, represented by \mZERO) symmetrically to both directions. (c) Illustration of the bifurcation mechanism for two atoms. The coaction of symmetric optical pumping (driving \mZERO~to \mUP~and \mDOWN) and spin-exchange collisions (allowing the pumping cycle to continue if the system arrives at \mDOWN\mUP~or \mUP\mDOWN) renders the maximally-polarized states (\mUP\mUP~and \mDOWN\mDOWN) the only basins of attraction. \label{fig:exp_system}}
\end{figure}

\begin{figure*}[th]    \centering
    \includegraphics[width=17.8 cm]{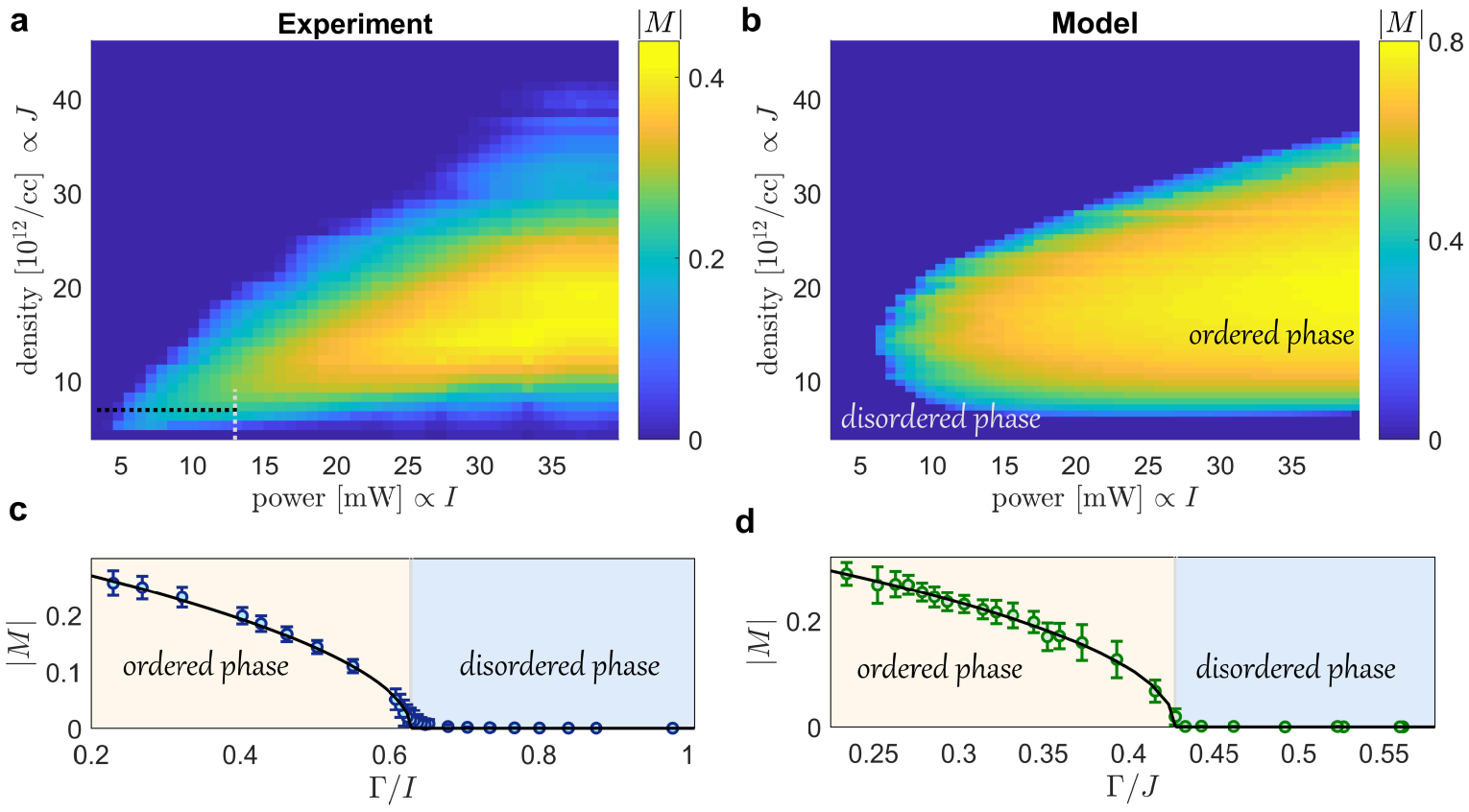}\caption{ \textbf{Magnetic phase diagram and power-law dependence.}  \textbf{a.} Measured steady absolute magnetization $|M|$ of the gas as a function of the atomic density and the power of the linearly-polarized pumping light. The atomic density sets the local spin-exchange collision rate $J$, while the pumping power sets the spin-alignment rate $I$. At moderate $J$ and elevated $I$, the spins gain order and align either parallel or anti-parallel to the magnetic field, manifesting a macroscopic ordered phase. The presented data are linearly interpolated; the raw data are shown on a scatter plot in Supplementary Figs.~\ref{fig:representations}(a, c). \textbf{b.} Simulated diagram using a nonlinear mean-field model (see Methods). \textbf{c-d.} Second-order phase transition, observed along the horizontal (black) and vertical (white) contours in \textbf{a}. Shown is measured magnetization as a function of  $\Gamma/I$ (\textbf{c}, with $J=3.8\Gamma$) and $\Gamma/J$ (\textbf{d}, with $I=4.5\Gamma$), where $\Gamma$ is the average spin-relaxation rate. We fit the data to a power law (black line) and find the critical exponents $\beta_I=0.53\pm0.04$ and  $\beta_J=0.49\pm0.02$ (See Methods).
    \label{fig:magnetic_phase_map}}
\end{figure*}

Here, we report on the observation of critical behavior of strongly interacting, warm cesium vapor. We measure the power-law dependence of the macroscopic magnetization on both the light intensity and gas density, as well as divergence of the susceptibility to an external spin imbalance. We identify these phenomena as a second-order magnetic phase transition and measure the phase diagram of the system. We observe divergence of the spin response time up to a few seconds when crossing the phase transition, two orders of magnitude longer than the $20$ millisecond spin lifetime. Furthermore, we observe a nine-fold improvement of the collective-spin lifetime near the phase boundary. Finally, we discuss interesting potential avenues for this accessible platform, including the exploration of new phases in gaseous systems, simulations of magnetic phenomena, and applications to quantum magnetic sensing.

\section{Results}
Figure \ref{fig:exp_system} presents the physical system. The cesium atoms, enclosed in a glass cell at near-ambient temperature, are unpolarized in the absence of optical fields. In this magnetically disordered phase, the cesium spins equally populate all 16 sub-states of the electronic ground level. To stimulate the transition into an ordered phase, we introduce linearly-polarized pumping light and increase the atomic density. The quantization axis $\hat{z}$ is set by an external magnetic field. We tune the optical frequency of the pumping light near the $D1$ optical transition from the lower hyperfine manifold ($F_\text{g}=3$) and set its polarization $\hat{x}$ perpendicular to $\hat{z}$. This configuration aligns the spins along $\hat{z}$ \cite{SP96}. It preferentially and symmetrically populates the two maximally-polarized states (with spin projection $m_{F}=\pm4$ along $\hat{z}$, in the upper hyperfine manifold $F_\text{g}=4$), marked by triangles in Fig.~\ref{fig:exp_system}(b). It does so at a rate $I$ linearly proportional to the intensity of the optical field (see Methods).

We maintain a constant temperature $T$, which we can vary in the range of  $55-120\,^{\circ}\text{C}$. The temperature sets the vapor pressure, originating from a reservoir (droplet) of cesium atoms, allowing us to control the atomic density and thus to determine the rate of spin-exchange collisions $J$. During a collision, the electronic spins of the two colliding atoms experience random, mutual precession, which conserves the total spin. The collisions change the internal atomic spin states, generate correlations between the atoms, and repopulate the lower hyperfine manifold. They comprise the microscopic inter-atomic interaction in our system necessary for the formation of an ordered phase. The interplay between optical pumping and spin-exchange collisions, leading to an alignment of the spins, is schematically illustrated in Fig.~\ref{fig:exp_system}c for the case of two atoms. It is important to note that the system does not reach equilibrium, but rather a non-equilibrium steady state, as there is a constant flow of energy from the pumping light through the system to the surrounding environment, which breaks the detailed balance condition. 

The cell also contains buffer gas that renders the atomic motion diffusive, yielding an average spin relaxation rate $\Gamma=58\,\text{s}^{-1}$ that is limited by collisions with the walls (see Methods). We monitor the macroscopic magnetization of the vapor $M$ using Faraday rotation measurements of off-resonant probe light \cite{happer2010optically, FaradayRot}. 

The cesium vapor becomes magnetized in the experiment for a range of collision rates $J$ and optical pumping (spin alignment) rates $I$. The spins then end up pointing either at the direction of the magnetic field ($M>0$) or opposite to it ($M<0$), with the sign varying randomly between experimental realizations. The measured absolute magnetization of the vapor $|M|$ at steady state is shown in Fig.~\ref{fig:magnetic_phase_map}a as a function of the pumping power and collision rate, which are respectively proportional to $I$ and $J$. We find a well-defined region in which the spins are ordered and $|M|>0$. We reach a  magnetization as high as $|M|=0.45$ in the ordered phase, where unity $|M|=1$ corresponds to maximal magnetization (all spins in the vapor maximally oriented along $\pm\hat{z}$). 

For other values of $I$ and $J$, the spins remain in a disordered phase with vanishing net magnetization in each realization. In Figs.~\ref{fig:magnetic_phase_map}c and \ref{fig:magnetic_phase_map}d, we present the magnetization as a function of $\Gamma/I$ and $\Gamma/J$ along two contours crossing the phase boundary (marked by dashed lines in Fig.~\ref{fig:magnetic_phase_map}a). We observe a critical dependence of the magnetization near the transition between the disordered and ordered phases. The continuous but sharp transition of the magnetization, which acts as an order parameter, indicates that the process is associated with a second-order phase transition. The data fits well to power-law functions with critical exponents $\beta_I=0.53\pm0.04$ and  $\beta_J=0.49\pm0.02$  (see Methods). 

One might expect that the magnetization only increases with the collision rate $J$. However, as clearly evident in Figs.\ref{fig:magnetic_phase_map}a, the magnetization decreases for large $J$, \textit{i.e.}, at elevated atomic densities. This is a result of the attenuation of the pumping light along the medium at high densities, which decreases the spatially-averaged spin-alignment rate below the critical value. Moreover, even when the spin-alignment rate is high, the critical behavior might be compromised by the large spatial inhomogeneity (along the $\hat{y}$ axis) due to this attenuation at elevated densities. Finally, at high pump powers, off-resonant excitation of the maximally-polarized states becomes a dominant relaxation mechanism of the magnetization. Together, these factors limit the parameter range at which we expect to observe the phase transition in our system. Nevertheless within a range of $\pm 20\%$ around the conditions highlighted in Fig.~\ref{fig:magnetic_phase_map}, we find that the critical exponents vary with standard deviations of $\sigma(\beta_I)=0.04$ and $\sigma(\beta_J)=0.06$. These are extracted from horizontal and vertical cuts of Fig.~S1e, which takes into account the attenuation of the spin-alignment rate at elevated densities.

\begin{figure}[t]    \centering
    \includegraphics[width=8.7 cm]{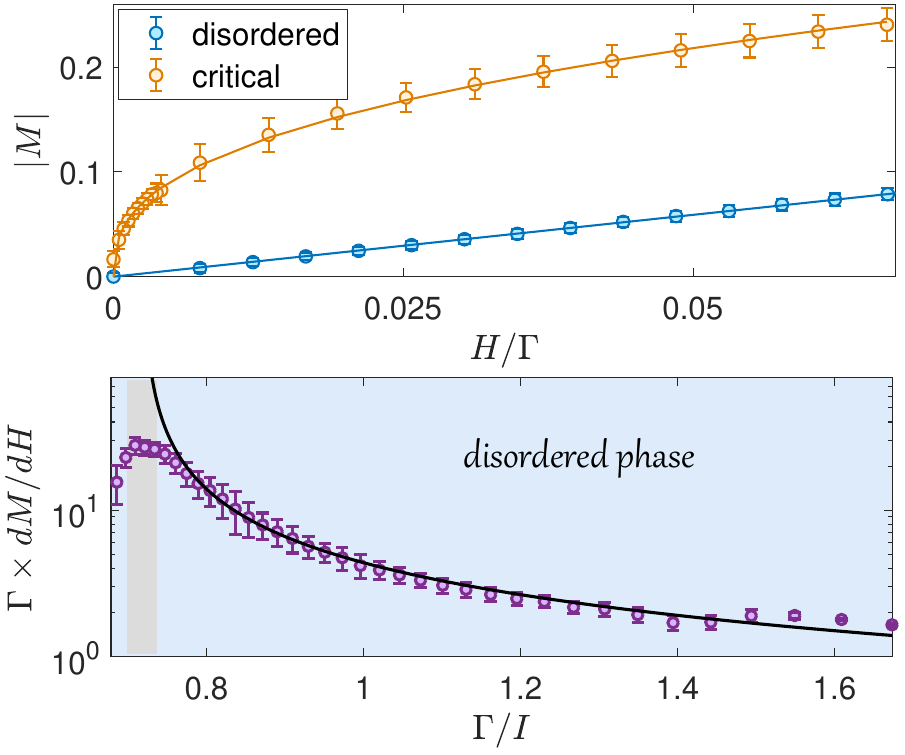}\setlength{\belowcaptionskip}{-4mm}\caption{\textbf{Critical behavior in the presence of a weak external bias}. \textbf{a.} The magnetization as a function of the bias strength, quantified by the optical-pumping rate $H$ induced by an auxiliary, circularly-polarized beam. The spin-exchange rate $J=2.3\Gamma$ is fixed. For $I=0$ (disordered phase) the magnetization increases linearly with the bias (blue circles), as expected for standard optical-pumping of spins. In contrast, for the critical value $I=1.5\Gamma$ at the phase boundary, the system becomes critical and the magnetization sharply increases (orange circles). Solid lines are fits to linear (blue) and power-law (orange) functions, the latter providing the critical exponent $\delta=2.65\pm0.09$. \textbf{b.} The susceptibility to an external bias $dM/dH$ (purple circles), diverging near the phase boundary (gray area). The fit to a divergent power-law (black line) provides the critical exponent $\gamma=0.94\pm0.10$ for the disordered phase. \label{fig:bias_beam} }
\end{figure}

To further explore the critical behavior near the boundary between the two phases, we measure the dependence of the magnetization on an external bias. The bias towards positive or negative $M$ is introduced by an auxiliary optical beam with circular polarization $\hat{\sigma}_+$ or $\hat{\sigma}_-$, respectively, see Fig.~\ref{fig:exp_system}a. The magnitude of the bias is given by the optical pumping rate $H$, which is linear in the bias beam's intensity. In Fig.~\ref{fig:bias_beam}a, we present the steady absolute magnetization as a function of $H/\Gamma$ near the phase boundary at $(I,J)=(1.5,2.3)\Gamma$ (orange circles) compared with the disordered phase at $(I,J)=(0,2.3)\Gamma$ (blue circles). In the disordered phase, the steady magnetization is determined by $M=H/(H+\Gamma)$ and, for a weak bias, grows linearly as $M\approx H/\Gamma$  (blue line). Importantly, the linear dependence of the magnetization on the bias beam's intensity is a universal property of weak resonant optical-pumping of uncorrelated atoms, independent of the transition strength or the particular atomic species. In contrast, near the phase boundary, we find a critical dependence of the magnetization on the bias beam, with a much sharper response. We fit the data to the power-law function $M=(H/\Gamma)^{1/\delta}$ and find the critical exponent $\delta=2.65\pm0.09$. The deviation from the standard optical-pumping relation indicates the emergence of correlations between the atoms.  

Next, we measure the susceptibility function $\chi= dM/dH$ near $H=0$, which determines the response of the system to a small external bias. Figure \ref{fig:bias_beam}b presents $\chi$ as a function of $I$ at $J=2.3$. We observe a striking increase of the susceptibility by more than an order of magnitude near the phase transition. Fitting $\chi$ to a divergent power-law function, we find the critical exponent $\gamma=0.94\pm0.10$ (see Methods). The power-law dependence of the magnetization and the divergence of the susceptibility near the phase boundary testify for the critical behavior of a magnetic, second-order phase transition. 

\begin{figure*}[t]
      \centering
    \includegraphics[width=17.8 cm]{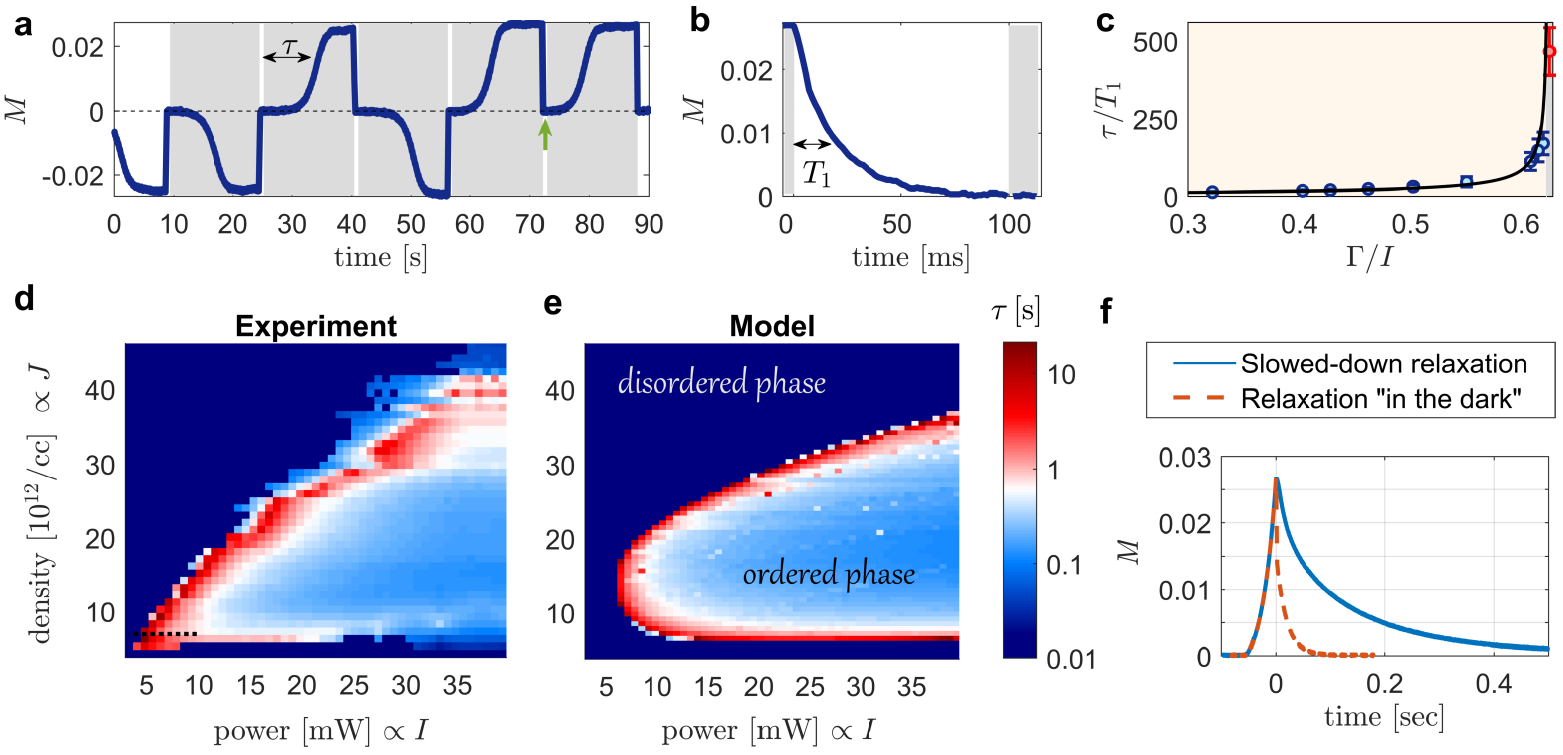}
    \caption{\textbf{Critical slow-down out of equilibrium.} \textbf{a.} Measured magnetization in response to optical-pumping pulses, with $I=1.6\Gamma$ and $J=3.7\Gamma$ close to the phase boundary. During the pumping on times (gray shaded areas), the magnetization builds up to $|M|\gtrsim 0.25$ in a random direction, at an average response time of $\tau=7.8\pm1.3$ seconds (time for the magnetization to reach $63\%$ of its steady value). \textbf{b.} During the off times, the system becomes disordered, and $|M|$ rapidly decays with the standard spin-lifetime $T_{1}=17$ ms. The data shown in \textbf{b} is marked by a green arrow in \textbf{a}. \textbf{c.} Divergence of the spin response time near the phase boundary and a fit to a power-law divergence (black line). Blue points are data obtained along the horizontal (dashed) contour in \textbf{d}, as a part of an automatic 40$\times$330 measurement, whereas the red point corresponds to the data in \textbf{a}, obtained by manually fine-tuning around the critical point. The bound $\tau\lesssim T_{1}$, valid for uncorrelated spins, is violated by a large factor. \textbf{d.} Measured spin response time in a logarithmic scale. The divergence appears at the phase boundary, where the spins are strongly correlated. The presented data is linearly interpolated; the raw data are shown in Supplementary Figs.~\ref{fig:representations}(b, d). \textbf{e.} Simulated spin-response time using a nonlinear mean-field model (see Methods) (f) Measured slow-down of spin relaxation near the phase transition. At $t=0$, we switch from a strong circularly-polarized pumping to a weak linearly-polarized pumping, the latter with an intensity slightly below its critical value. The relaxation is substantially slower (solid blue line) than the standard relaxation in the dark (dashed orange line). These preliminary data indicate a 9-fold slowing down of the collective spin relaxation.
    \label{fig:time_phase_map}}
    \end{figure*}

Second-order phase transitions exhibit critical, scale-invariant behavior even out of equilibrium. Here we explore the dynamical transition from a disordered to an ordered phase by temporally varying $I$ in a pulsed waveform. We add no bias and use a fixed $J=3.7\Gamma$, for which the critical value of $I$ is $1.6\Gamma$. For $I>1.6\Gamma$, a magnetization initially at $M=0$ builds up to a finite steady value, randomly in either of the two directions. Figure \ref{fig:time_phase_map}a shows an example of the measured magnetization subject to a periodic sequence of pumping pulses (gray areas) tuned slightly above $I=1.6\Gamma$.
We observe a slowdown of the polarization process, with an average response time of $\tau=7.8\pm1.3$ seconds. Between the pumping pulses (white areas), the magnetization rapidly vanishes, with a decay rate corresponding to the spin lifetime in the dark ($T_{1}=17$~ms at that temperature), as shown in Fig.~\ref{fig:time_phase_map}b. Importantly, as absorption of photons contributes to the spin relaxation, the response time of uncorrelated spins by resonant optical pumping is always shorter than their lifetime in the dark $T_1$. In contrast, here near the phase boundary, $\tau$ becomes longer than $T_1$ by a remarkable factor of 460. The observed response time varies between the pulses due to the critical nature of the phenomenon, and nevertheless, it is always at least two orders of magnitude larger than $T_1$.

The slow down of the spin response time is a critical phenomenon associated with the phase transition. Figure \ref{fig:time_phase_map}c shows the dependence of $\tau$ on $I$ along the contour $J=3.7\Gamma$ crossing the phase boundary (dashed line in Fig.~\ref{fig:magnetic_phase_map}a and  Fig.~\ref{fig:time_phase_map}d). 
We observe a divergence of $\tau$ near $I=1.6\Gamma$ and find the critical exponent $z\nu=0.86\pm0.07$ by fitting to a divergent power-law function (see Methods). The critical exponent varies with a standard deviation of $\sigma(z\nu)=0.13$ within a range of $\pm 20\%$ around $J=3.7\Gamma$. The spin response time diverges at the entire phase boundary, as shown in Fig.~\ref{fig:time_phase_map}d. The observation of a critical divergence of the response time (Fig.~\ref{fig:time_phase_map}d) near the boundary of the ordered phase (Fig.~\ref{fig:magnetic_phase_map}a) attests to the collective nature of the phase transition. 

Finally, in addition to the elongation of the spin buildup time discussed above, one expects near the phase transition boundary an elongation of the spin relaxation time. We explore this by first initializing the spins in the ordered phase (using a strong circularly-polarized beam) and subsequently monitoring the spin relaxation in the presence of a linearly-polarized light slightly below the phase transition, \textit{i.e.,} in the disordered-phase region near the phase boundary. We compare the relaxation near the phase boundary (Fig.~\ref{fig:time_phase_map}f, solid blue line) to the standard relaxation away from the boundary (Fig.~\ref{fig:time_phase_map}f, dashed orange line). Fitting the data to a decaying exponent, we find that the slowed-down relaxation time $\tau=165\pm14$~ms is longer by a factor of nine than the lifetime in the dark $T_1=18.3\pm0.1$~ms at the same temperature.

To model the observed phenomena, we employ a nonlinear mean-field model that describes the evolution of the density matrix of the mean cesium spin in the vapor. The theoretical model, described in Methods, captures the main features of the emergent phase transition, including the bi-stability, the power-law dependencies, and the divergence of the susceptibility and the spin response time. In Figs.~\ref{fig:magnetic_phase_map}b and \ref{fig:time_phase_map}e, we present the numerically calculated phase diagram and the spin response time, which agree with the measured results reasonably well.

\section{Discussion}
In summary, we observe a magnetic phase transition of warm cesium atoms stimulated by linearly-polarized light and by frequent spin-exchange collisions. We characterize the phase diagram, reveal a critical power-law behavior near the phase transition, and observe a critical slow-down of the spin response time. The substantial deviation of our observations from standard experiments, in which the atoms are uncorrelated, testifies for the collective nature of the ordered phase and the transition. 

This work opens new avenues for studying correlated phases in gaseous systems at ambient conditions. First, it is remarkable that a simple optical driving affecting each spin independently renders the inter-particle interactions dominant over the gas entropy, resulting in an emerging correlated phase. It would be interesting to explore optical-driving techniques that can enhance this dominance further in various atomic and molecular gases, potentially resulting in correlated phases at elevated temperatures. Engineering the range of interactions between the spins is a second avenue, which could lead to nontrivial spatial correlations in the gas and potentially to the formation of spatial domains, \textit{e.g.}, via the Kibble-Zurek mechanism. The critical exponents we measured fall into the universality class of mean-field models \cite{Mermin502} (cf.~Table~\ref{table:1}) due to rapid diffusion of the atoms. In our system, the atoms diffuse on a short time scale ($\sim T_{1}$), thus enabling spin-exchange interaction between initially distant atoms and effectively rendering the range of the interaction long. Slowing down the diffusion (\textit{e.g.},~by increasing the buffer-gas pressure) can reduce the interaction range. A third avenue can exploit the long spin coherence of the gas to explore other possible nonequilibrium phases. For example, a periodic variation of the magnetic field combined with the effectively long-range interaction between the spins can potentially make the system resilient to disorder (\textit{e.g.} in the magnetic field or the beam intensity) and drive the system into a time-crystalline phase.

The new platform can potentially be utilized in applications such as sensing and spin simulators. Warm atomic spins are useful sensors as they strongly couple to light and have long coherence times (up to hundreds of milliseconds) and a large number of particles at or above room-temperature \cite{Kitching2011}. These sensors often measure the response of the spins to small external fields using optical means. Divergence of the spins' response to external fields near the phase boundary can considerably enhance the signal over the noise up to the standard quantum limit (which is typically hard to reach). For example, the critical behaviour observed in Fig.~\ref{fig:bias_beam} enables enhanced detection of the optical polarization and circularity of the beam with respect to a standard measurement in a disordered phase. Furthermore, longer spin lifetimes often improve the performance of spin sensors, both in transient and in steady-state operation \cite{RevModPhys.89.035002}. Therefore, the elongation of the spin lifetime near the phase boundary, which is optically controllable, can potentially improve sensing performance.

Simulation of complex many-body phenomena and computation of optimization problems utilize various platforms, such as ultracold atomic gases and coupled laser arrays. These simulators consist of a network of spins, whose evolution under a controllable interaction Hamiltonian is studied. The new platform can be employed in several configurations for constructing efficient networks in a relatively simple setup at room temperature. One configuration can use a large but thin vapor cell, illuminated by a wide beam whose optical polarization is spatially modulated. In such a network, spin-exchange and diffusion lead to interaction between neighboring domains, each initialized in a correlated ordered phase, whereas the light modulation enables local engineering of the desired interaction Hamiltonian. Other configurations can use arrays of miniature cells, each acting as a single computational spin, and an array of optical beams for the network linkage, exploiting the strong bi-directional coupling between spins and light which is enhanced near the phase boundary. The long spin lifetimes of this platform could be particularly interesting in simulations of out-of-equilibrium phenomena such as anomalous thermal relaxations in magnetic systems \cite{gal2020precooling,yang2020non,klich2019mpemba,lu2017nonequilibrium,baity2019mpemba} as well as in computation of a wide class of optimization problems \cite{inagaki2016coherent,mcmahon2016fully,takata201616}.

\begin{table}[b]
\centering
\begin{tabular}{|c|c|c|c|}
 \hline
 Exponent & Measured value & Mean-field value & Relation \\ [0.5ex] 
 \hline \hline
$\beta_{I}$ & $0.53\pm0.06$ & 0.5 & $M\propto I^{-\beta_{I}}$\\ 
 \hline
$\beta_{J}$ & $0.49\pm0.06$ & 0.5 & $M\propto J^{-\beta_{J}}$\\ 
 \hline
 $\delta$ & $2.65\pm0.09$ & 3 & $M\propto H^{1/\delta}$\\
 \hline
 $\gamma$ & $0.94\pm0.10$ & 1 & $\nicefrac{dM}{dH}\propto I^{\gamma}$ \\
 \hline
 $z\nu$ & $0.86\pm0.15$ & 1 & $\tau\propto I^{-z\nu}$ \\
 \hline
\end{tabular}
\caption{Measured critical exponents compared with the exponents of the mean-field universality class. Simplified power-law relations near the phase boundary are presented for context, with $M$ denoting the magnetization, $I$ the optical spin-alignment rate, $J$ the spin-exchange collision rate, $H$ the bias pumping rate, and $\tau$ the dynamical response time. Full expressions of the critical functions are given in Methods. The fits of the power-law functions in a logarithmic scale is presented in Supplementary Fig.~\ref{fig:averaged_I}. The noted errors include the fitting uncertainty to the power law and, for $\beta_I$, $\beta_J$, and $z\nu$, also the standard deviation of the mean values within the analyzed parameter range.}
\label{table:1}
\end{table}

\begin{acknowledgments}
We acknowledge financial support by the estate of Emile Mimran, the Israel Science Foundation, the US-Israel Binational Science Foundation (BSF) and US National Science Foundation (NSF), the Minerva Foundation with funding from the Federal German Ministry for Education and Research, the Shlomo and Michla Tomarin career development chair, the Abramson Family Center for Young Scientists, and the Laboratory in Memory of Leon and Blacky Broder. This research was facilitated by the Talpiot Program 18-19-30-36.
\end{acknowledgments}

\clearpage
\newpage

\appendix

\part*{$\;\;\;\;\;\;\;\;\;\;\;\;\;\;$Methods}

\subsection*{Detailed experimental setup}
We use a cubic borosilicate glass cell of length $L=15$ mm containing cesium vapor, and a buffer-gas mixture of $\text{N}_2$ and neon, $16.5$ Torr each. These gases slow down the diffusion of cesium atoms to the cell walls, and the $\text{N}_2$ additionally enables the non-radiative decay (quenching) of electronically-excited cesium atoms. The buffer gases broaden the cesium optical lines to $\gamma_{\text{c}}=137\,\text{MHz}$ (HWHM), yet the four optical transitions $F_\text{g}=\{3,4\}\rightarrow F_\text{e}=\{3,4\}$ of the $D_{1}$ line remain resolved. We heat the cell using high-frequency electrical current at 390 kHz flowing through high-resistance twisted-pair wires in a custom oven. We control the magnetic field in the cell using three pairs of Helmholtz coils and set a constant magnetic field of $B=1\,\text{G}$ along $\hat{z}$. The coils are located within four $\mu$-metal layers, shielding the cell from external magnetic fields.

The linearly-polarized pumping beam originates from a free-running DBR diode laser at 895 nm. The laser frequency is blue-detuned by $\Delta=700\,\text{MHz}$ from the $F_\text{g}=3\rightarrow F_\text{e}=4$ optical transition. We control the beam power using a commercial noise-eater consisting of a liquid crystal, polarizing beam splitter, and a photodetector. The beam then passes through a mechanical shutter and a high-quality linear polarizer, which sets the polarization of the beam to be linear along $\hat{x}$, and finally through a $\lambda/4$ waveplate mounted on a precision, computer-controlled rotating mount. For most of the experiments, the fast axis of the waveplate is carefully aligned with the direction of the linear polarization, unaffecting the linear polarization of light. Nevertheless, when needed, rotation of the waveplate enables rapid calibration of the maximal magnetization within the experimental sequence (as detailed below). The Gaussian beam is then expanded to a $1/e^2$ radius of $1$ cm to cover the entire cell area and enter the cell in the $\hat{y}$ direction. We measure the magnetization in a set of $330\times40$ measurements varying the vapor density in the range of  $7\times10^{11}\,\text{cm}^{-3}\leq n\leq5\times10^{13}\,\text{cm}^{-3}$  and the optical intensity of the pumping field in the range of  $2\,\text{mW}\leq\Phi\leq40\,\text{mW}$ . 

We monitor the magnetization using a $\hat{y}$-polarized, $3$-mW probe beam that propagates along $\hat{z}$. The beam originates from another free-running DBR diode laser at 895 nm. It has a Gaussian profile with $1/e^2$ radius of $1.5$ cm, covering much of the atoms in the cell. The beam is blue-detuned by $100\,\text{GHz}$ from the $D_{1}$ lines to avoid photon absorption. It probes the $z$-component of the electron spin of the gas via Faraday rotation. The beam then goes into a balanced polarimetry setup, which outputs a signal proportional to the magnetization along $\hat{z}$.

The bias beam originates from a third free-running DBR diode-laser at 895 nm, whose power is controlled by an acusto-optic modulator and a commercial intensity noise-eater. 
A $\lambda/4$ waveplate renders the polarization of the beam circular, and it is combined with the probe beam using a non-polarizing beam splitter,
as shown in Fig.~\ref{fig:exp_system}. The beam has a Gaussian profile with $1/e^2$ radius of $1$ cm. It is 1.2-GHz blue-detuned from the $F_\text{g}=3\rightarrow F_\text{e}=4$ transition for the critical behavior experiments, and it is set to resonance with the $F_\text{g}=4\rightarrow F_\text{e}=3$ for calibration experiments.  

Our measurements are found sensitive both to the detuning of the pumping beam and to the magnitude of the magnetic field. Drift in the detuning predominantly affects the rate $I$, while drift in the magnetic field was found to vary the spin response time. Therefore, both quantities are monitored and kept constant during the experiment.

\subsection*{Experimental calibrations}
The spin lifetime $T_{1}=1/\Gamma$ is determined by measuring the decay rate of spins oriented along the magnetic field in the absence of resonant optical fields (measurement in the dark). In this measurement, the spins are first optically pumped by two circularly polarized beams resonant with the  $F_\text{g}=3\rightarrow F_\text{e}=4$ and  $F_\text{g}=4\rightarrow F_\text{e}=3$ transitions along $\hat{y}$, then rotated to the $\hat{z}$ axis by a magnetic field pulse, and finally measured with the off-resonant probe while the pumping beams are off. The relaxation rate has a small, linear dependence on the cell's temperature, satisfying $\Gamma(T)=\Gamma_{0}+0.35(T \,[^{\circ}\text{C}]-75)$ in the tested range ($55-120\,^{\circ}\text{C}$), predominantly due to the temperature dependence of the diffusion coefficient of the cesium atoms. In the entire analysis and figures, we use the constant value $\Gamma=\Gamma_{0}=58\,\text{s}^{-1}$. Furthermore, in the measured and simulated response-time data (Figs.~\ref{fig:time_phase_map}d,e), we set $\tau=\text{T}_1$ for all measurements where the spin response was smaller than $10^{-3}$ of the maximal measured magnetization.

The spin-alignment rate $I$ is independently determined by measuring the excess decay rate of the spins in the presence of the pumping beam at $T=75 ^\circ\text{C}$. We find a linear dependence on the intensity of the beam $\Phi$, with a ratio $I/\Phi=457\,\text{cm}^{2}/\text{J}$. Note that the local spin-alignment rate varies across the profile of the pumping beam, and therefore we always refer to an average rate across the beam. The pumping beam is also attenuated along the propagation direction ($\hat{y}$) due to absorption by the atoms. The $y$-dependent spin-alignment rate is therefore given by $I\exp(-n\sigma_{\text{e}} y)$ where $n(T)$ is the vapor density and $\sigma_{\text{e}}(\Delta)$ is the absorption cross-section of the beam. 

The spin-exchange rate $J(T)$ is independently determined by a measurement of the relaxation of the spins transverse to the magnetic field, in the absence of resonant optical fields. In this measurement, we weakly pump the spins along $\hat{y}$, apply a magnetic field $B$ along $\hat{x}$, and monitor the precession of the spins, decaying at a decoherence rate $\Gamma_2(B)$. At each temperature, we determine the spin-exchange rate  $J=[\Gamma_2(B)-\Gamma_2(0)]/q$ from the measured relaxation at high magnetic field $B=2\,\text{G}$ and by subtracting the effect of other field-independent relaxations $\Gamma_2(0)$ \cite{serf1977, katz2013nonlinear}. Here $q=4.57$ is the numerical slow-down factor that accounts for the reduction of the rate by coupling to the nuclear spin $I=7/2$ \cite{serf1977}. 

The bias rate $H$ is independently determined by measuring the magnetization as a function of the intensity $\Phi_\text{H}$ of the bias (circularly-polarized) beam. We fit the measured magnetization to the function $M=a\Phi_\text{H}/(a\Phi_\text{H}+\Gamma)$ with $a=1.179\pm0.005$ and determine the linear coefficient $H/\Phi_\text{H}=99\,\text{cm}^{2}/\text{J}$ . 

As we increase the temperature of the cell to increase $J$, the vapor density, and therefore the number of gaseous spins in the cell, increase as well. We thus calibrate for the maximal polarization of the vapor at each temperature to properly determine the magnetization $M$, which describes the portion of polarized spins in the gas. For that, we use two strong circularly polarized beams that cover the entire cell and optically pump the spins along  $\hat{y}$. We then turn the beams off and apply a magnetic field along $B_{x}$ that stimulates the precession of the spins. The precession amplitude corresponds to the maximal signal obtained by our probe beam, which is identified as the maximal polarization and used to calibrate the magnetization in the experiments.

The residual circularity of the polarization of the pumping field is automatically zeroed within each experimental sequence by applying the same technique as in the spin-exchange rate $J(T)$  calibration for varying $\lambda/4$ waveplate angles around the known optimal point. Since the waveplate is mounted on a precision, computer-controlled rotating mount, we automatically repeat this process until the procession amplitude is minimal. This ensures the beam has a minimal circular polarization. 

\subsection*{Critical behavior}
Near the phase boundary, the data exhibit power-law dependence with critical exponents. Here we describe the fitting procedure we use to determine these exponents. 

We determine the critical exponent $\beta_{I}$ by fitting the data in Fig.~\ref{fig:magnetic_phase_map}c to the function $M(I>I_0)=M_{0}(1-I_{0}/I)^{\beta_{I}}$  and $M(I<I_0)=0$. For proper fitting, we first estimate the initial guess for $I_{0}$ and $\beta_{I}$ by fixing the value of one parameter and fitting for the other and use those results for the final fit. For $J=3.8\Gamma$ (black horizontal dashed line in Fig.~\ref{fig:magnetic_phase_map}a), we find the critical exponent $\beta_I=0.53\pm0.04$ and the critical spin-alignment rate $I_{0}=1.639\pm0.006\Gamma$. 
Similarly, we determine the critical exponent $\beta_{J}$ by fitting the data in Fig.~\ref{fig:magnetic_phase_map}d to the function $M(J>J_0)=M_{0}(1-J_{0}/J)^{\beta_{J}}$  and $M(J<J_0)=0$. For $I=4.5\Gamma$ (white vertical dashed line in Fig.~\ref{fig:magnetic_phase_map}a), We find the critical exponent $\beta_J=0.49\pm0.02$ and the critical spin-exchange rate $J_{0}=(2.393\pm0.004)\Gamma$.

We determine the critical exponent $\gamma$ by fitting the measured susceptibility in Fig.~\ref{fig:bias_beam}b to the function $\chi=\chi_{0}(I_0/I-1)^{-\gamma}$ in the disordered phase for $I<I_0$. As with the critical exponent $\beta$ fitting, we first estimate the initial guess for $I_0$ and $\gamma$ by fixing the value of one parameter and fitting for the other and use those results for the final fit. We also exclude points that are close to the maximal measured one and use weights $w=(\Gamma/I)^3$ in order to compensate for the finite values of the data around the critical point compared with the divergent values of the model. For $J=2.3 \Gamma$, we find the critical exponent $\gamma=0.94\pm0.10$ and the critical spin-alignment rate $I_0=1.393\pm0.036\Gamma$ in the disordered phase.

We determine the critical exponent $z\nu$ by fitting the spin response time $\tau$ in Fig.~\ref{fig:time_phase_map}c to the power-law function $\tau=\tau_{0}(1-I_{0}/I)^{-z\nu}$. 
We fit the functions to 11 measurements in the range of exchange rates $J=3.47-3.96\,\Gamma$  (equivalent to $\text{T}=87\pm1\,^{\circ}\text{C}$) in order to comprehend the deviation of the model parameters. As before, we use a three-step fitting scheme, exclude points that are close to the maximal measured one, and use weights. We then find the critical exponent $z\nu=0.86\pm0.07$ and the critical spin-alignment rate $I_0=1.599\pm0.004\Gamma$.

\subsection*{Spin-alignment by linearly polarized light}
The linearly-polarized pumping light resonantly interacts with the optical transition $F_\text{g}=3\rightarrow F_\text{e}=4$. As the electric field is perpendicular to the magnetic field, upon absorption of a photon, the atom is excited and its spin projection along the magnetic field $m_{F}$ changes by either $+1$ or $-1$. Importantly, this interaction preferably increases the absolute spin projection $|m_{F}|$ \cite{SP87}. An atom with $m_{F}>0$ would preferably be excited while increasing its spin to $m_{F}+1$, and, symmetrically, an atom with $m_{F}<0$ would preferably decrease its spin to $m_{F}-1$. This preference is quantified in Table \ref{table:2}, which presents the probability for changing the ground-level spin projection by absorption of a linearly-polarized photon. The on-average increase of $|m_{F}|$  drives the spins towards either $m_{F}=4$ or $m_{F}=-4$, thus generating symmetric alignment.

\begin{table}[h!]
\centering
\begin{tabular}{||c c c c||}
 \hline
 $|m_F|$ & $~p_{|m_{F}|\rightarrow |m_{F}|+1}~$ & $~p_{|m_{F}|\rightarrow |m_{F}|-1}~$ & $\Delta p$ \\ [0.5ex] 
 \hline\hline
 0 & 1/2 & 1/2 & 0 \\ 
 \hline
 1 & 15/21 & 6/21 & 9/21 \\
 \hline
 2 & 7/8 & 1/8 & 6/8 \\
 \hline
 3 & 28/29 & 1/29 & 27/29 \\
 \hline
\end{tabular}
\caption{Transition probabilities for changing the spin projection $|m_F|$ of cesium atoms in $F_\text{g}=3$ upon absorption of the pumping beam. The pumping tends to increase $|m_F|$, as indicated by the positive $\Delta p=p_{|m_{F}|\rightarrow |m_{F}|+1}-p_{|m_{F}|\rightarrow |m_{F}|-1}$, thus driving the spins towards the maximally-polarized states $m_{F}=\pm4$  in $F_\text{g}=4$.}
\label{table:2}
\end{table}

\subsection*{Theoretical model}
We implement a mean-field model describing the dynamics of the mean density matrix $\bar{\rho}$ of a single atom in the vapor, following the model by Happer \textit{et al.}~\cite{happer2010optically}. The density matrix 
\begin{align}\bar{\rho}=\left(\begin{array}{cc}
\bar{\rho}_\text{e} & \bar{\rho}_\text{eg}\\
\bar{\rho}_\text{ge} & \bar{\rho}_\text{g}
\end{array}\right)\end{align}
consists of the 16 spin state in the ground level $F_\text{g}=\{3,4\}$ denoted by $\bar{\rho}_\text{g}$, the 16 spin states in the excited level $F_\text{e}=\{3,4\}$ denoted by $\bar{\rho}_\text{e}$, and the optical-coherence matrices between the two $\bar{\rho}_\text{ge}$ and $\bar{\rho}_\text{eg}$. We describe the evolution of the density matrix by solving the non-linear Liouville equation 
\begin{align} \label{eq:1}
\partial_t{\bar{\rho}} = &-\frac{i}{\hbar}\bigl[H + V_{\text{L}}, \bar{\rho}\bigr] +\mathcal{L}(\bar{\rho}).\end{align}
$H=H_\text{g}+H_\text{e}$ is the spin Hamiltonian of an alkali atom. Both ground ($H_\text{g}$) and excited-level ($H_\text{e}$) Hamiltonians consist of the hyperfine interaction $A\mathbf{I}\cdot\mathbf{S}$ and the interaction with a magnetic field, predominantly of the electron spin $gB\cdot \textbf{S}$. Here $\mathbf{S}$ denotes the electronic spin operator and $\mathbf{I}$ denotes the nuclear spin operator. We use $A=2.3\,\text{GHz}$ and $g=2.8\,\text{MHz}/\text{G}$ in $H_\text{g}$, and we use $A=290\,\text{MHz}$ and $g=0.9\,\text{MHz}/\text{G}$ in $H_\text{e}$. The second term in Eq.~(\ref{eq:1}) describes the atom-photon interaction $V_{\text{L}}=-\mathbf{E}\cdot\mathbf{D}$, coupling the oscillating electric field of the optical pumping field $\mathbf{E}=\textbf{E}_{0}e^{i(\textbf{k}\cdot\textbf{v}-\omega\,t)}$ with the atomic dipole operator $\mathbf{D}$. The term $\textbf{k}\cdot\textbf{v}$ denotes the Doppler shift due to the finite velocity $v$ of that atom. The third term $\mathcal{L}(\bar{\rho})$ describes the coupling of the spins to other degrees of freedom via radiative or collisional channels in the ground $\mathcal{L}_\text{g}(\bar{\rho})$ and excited levels $\mathcal{L}_\text{e}(\bar{\rho})$, as well as dephasing of the optical coherences at a rate $\gamma_\text{c}$. 

Rapid collisions with buffer-gas atoms increase $\gamma_\text{c}$ significantly (pressure broadening) with respect to both the spontaneous emission rate and the Rabi frequency of the optical pumping fields. In a frame rotating near the light frequency at a detuning $\Delta$, the optical coherence $\bar{\rho}_\text{eg}$ is hence maintained in a quasi-steady state, satisfying $\bar{\rho}_\text{eg} =w\bar{\rho}_\text{g}-\bar{\rho}_\text{e}w$. The operator $w=\bigl\langle\mathcal{E}^{-1}(\textbf{E}_{0}\cdot \textbf{D})\bigr\rangle_{v}$ denotes the fraction of optical coherence, with $\mathcal{E}=H_\text{e}-H_\text{g}+\hbar(\textbf{k}\cdot\textbf{v}-\Delta-i\gamma_\text{c})$ accounting for the optical detuning and dephasing of the different transitions. The operation $\bigl\langle\bigr\rangle_{v}$ denotes thermal averaging using the Maxwell-Boltzmann distribution, to properly account for Doppler broadening of the optical transition. 

Atoms in the excited level experience rapid relaxation, which we model by
\begin{align} 
\mathcal{L}_\text{e}(\bar{\rho})=-\gamma_\text{q}\bar{\rho}_\text{e}-\gamma_\text{p}\bigl(\tfrac{3}{4}\bar{\rho}_\text{e} -\textbf{S}\bar{\rho}_\text{e}\textbf{S}\bigr)\label{eq:2}\end{align}
The first term denotes de-excitation (quenching) of the population from the excited level to the ground level at a rate $\gamma_\text{q}$, including both spontaneous emission and collisions with $\text{N}_2$ molecules. The second term describes the destruction of the electron spin in the excited-level manifold by collisions with buffer gas atoms at a rate $\gamma_\text{p}$. As an approximation, our numerical calculation assumes a quasi-steady-state solution of $\bar{\rho}_\text{e}$ in Eq.~(\ref{eq:1}), which adiabatically follows the dynamics of the ground-level density matrix $\bar{\rho}_\text{g}$. We then numerically solve only the dynamics of the density matrix $\bar{\rho}_\text{g}$.

The dynamics of the ground-level density matrix $\bar{\rho}_\text{g}$ is modeled by 
\begin{align}\label{eq:3}\mathcal{L}_\text{g}(\bar{\rho}) & =\tfrac{2\gamma_\text{q}}{3D^{2}}\textbf{D}^{\dagger}\bar{\rho}_\text{e}\textbf{D}-\Gamma\bigl(\tfrac{3}{4}\bar{\rho}_\text{g}-\textbf{F}\bar{\rho}_\text{g}\textbf{F}\bigr).\\
 & -qJ\left[\tfrac{3}{4}\bar{\rho}_\text{g}-\textbf{S}\bar{\rho}_\text{g}\textbf{S}+\textbf{M}\cdot(\bar{\rho}_\text{g}\mathbf{S}+\mathbf{S}\bar{\rho}_\text{g}-2i\mathbf{S}\times\bar{\rho}_\text{g}\mathbf{S})\right]\nonumber&\end{align}
The first term describes the repopulation of the ground level by quenched excited atoms, where $D$ is the amplitude of the dipole moment of the optical transition. The second term describes the destruction of the total spin $\textbf{F}=\textbf{I}+\textbf{S}$ at a rate $\Gamma$, predominantly due to diffusion to the cell walls. The last term describes the effect of spin-exchange collisions, affecting the electron spin at a rate $qJ$. Importantly, the spin-exchange interaction has a nonlinear (quadratic) dependence on the density matrix, since the magnetization of the vapor is given by $\textbf{M}=\text{Tr}(\bar{\rho}_\text{g}\textbf{S})$. The quadratic dependence enables the emergence of bi-stable steady solutions and manifests the correlations induced by different atoms via the exchange interaction. It is therefore crucial for the emergence of the ordered phase and the observed critical phenomena.

We numerically calculate the steady state and the buildup of the magnetization of the vapor by solving Eqs.~(\ref{eq:1}-\ref{eq:3}) using an initially unpolarized state. We use $\gamma_\text{c}=1.86\,\text{GHz}$, $\gamma_p=219\,\text{MHz}$,  $\gamma_q=265\,\text{MHz}$,  $q=4.57$, $\Delta=700\,\text{MHz}$, and $\Gamma(T)=\Gamma_{0}+0.35(T \,[^{\circ}\text{C}]-75)$ to model the parameters of the experiment \cite{happer2010optically}. We run the simulation for a range of 70 vapor densities $7\times10^{11}\,\text{cm}^{-3}\leq n\leq5\times10^{13}\,\text{cm}^{-3}$ and 270 optical intensities $2\,\text{mW}\leq\Phi\leq40\,\text{mW}$. In terms of the model parameters, the effective rates used in the main text are given by $J=n\langle\sigma_{\text{ex}}v\rangle_{v}$ and  $I=s\,\text{E}_0^2\,\exp(-\text{OD})$. Here $\langle\sigma_{\text{ex}}v\rangle_{v}=7\times10^{-10}\,\text{cm}^3\,\text{s}^{-1}$, $s=220\,\text{MHz}\,(\text{mW}/{\text{cm}}^2)^{-1}$, and the optical depth $\text{OD}=n\sigma_{\text{e}} L$ is given by the measured value at each vapor density.

To alleviate the numerical complexity of the model, we zero the rapidly oscillating hyperfine coherences of $\bar{\rho}_\text{g}$  at every step of the simulation. This approximation corresponds to the rapid decay of these coherences due to spin-exchange relaxation in the experiment.
In the simulations presented in Fig.~\ref{fig:magnetic_phase_map}b and Fig.~\ref{fig:time_phase_map}e, we further zero the Zeeman coherences of $\bar{\rho}_\text{g}$ at every step of the simulation to speed up the calculation. Indeed in practice, spin-exchange collisions at a large magnetic field partially relax these coherences. We demonstrate in Supplementary Fig.~\ref{fig:magnetic_field} that the effect of these coherences is limited to variations of the spin response time at the phase boundary at lower densities.

\bibliographystyle{apsrev4-2}
\bibliography{main}

\appendix 
\setcounter{equation}{0}
\setcounter{figure}{0}
\setcounter{table}{0}
\makeatletter
\renewcommand{\theequation}{S\arabic{equation}}
\renewcommand{\thefigure}{S\arabic{figure}}
\renewcommand{\citenumfont}[1]{S#1} 

\newpage

\begin{figure*}[bt] 
\centering
\includegraphics[width=16.6 cm]{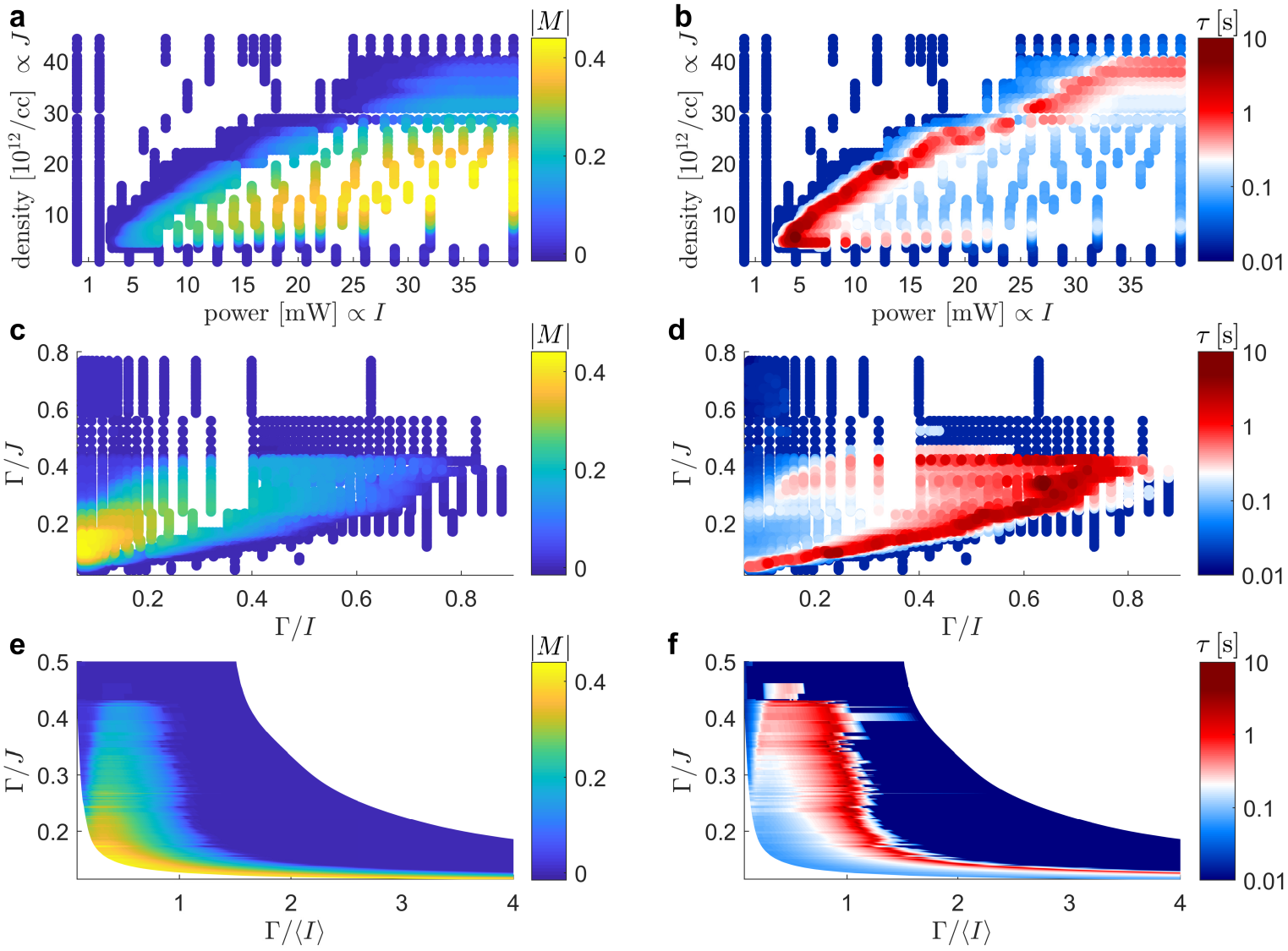}\setlength{\belowcaptionskip}{-4mm}\caption{\textbf{Different representations of the phase-diagram data}. \textbf{a-b.} Raw data of the measured steady magnetization $|M|$ and spin-response time $\tau$. \textbf{c-d.} Raw data of $|M|$ and $\tau$ as a function of the normalized inverse rates $\Gamma/J$ and $\Gamma/I$. \textbf{e-f.} Interpolated $|M|$ and $\tau$ data as a function of $\Gamma/J$ and $\Gamma/\langle I \rangle$, where the spatially-averaged optical spin-alignment rate is given by $\langle I \rangle$=$\int_{0}^{L} I\exp(-n\sigma_{\text{e}} y) dy$, with $n(T)$ the vapor density and $\sigma_{\text{e}}(\Delta)$ the absorption cross-section for the pumping field.\label{fig:representations} } 
\end{figure*}

\begin{figure*}[bt]    \centering
    \includegraphics[width=16.6 cm]{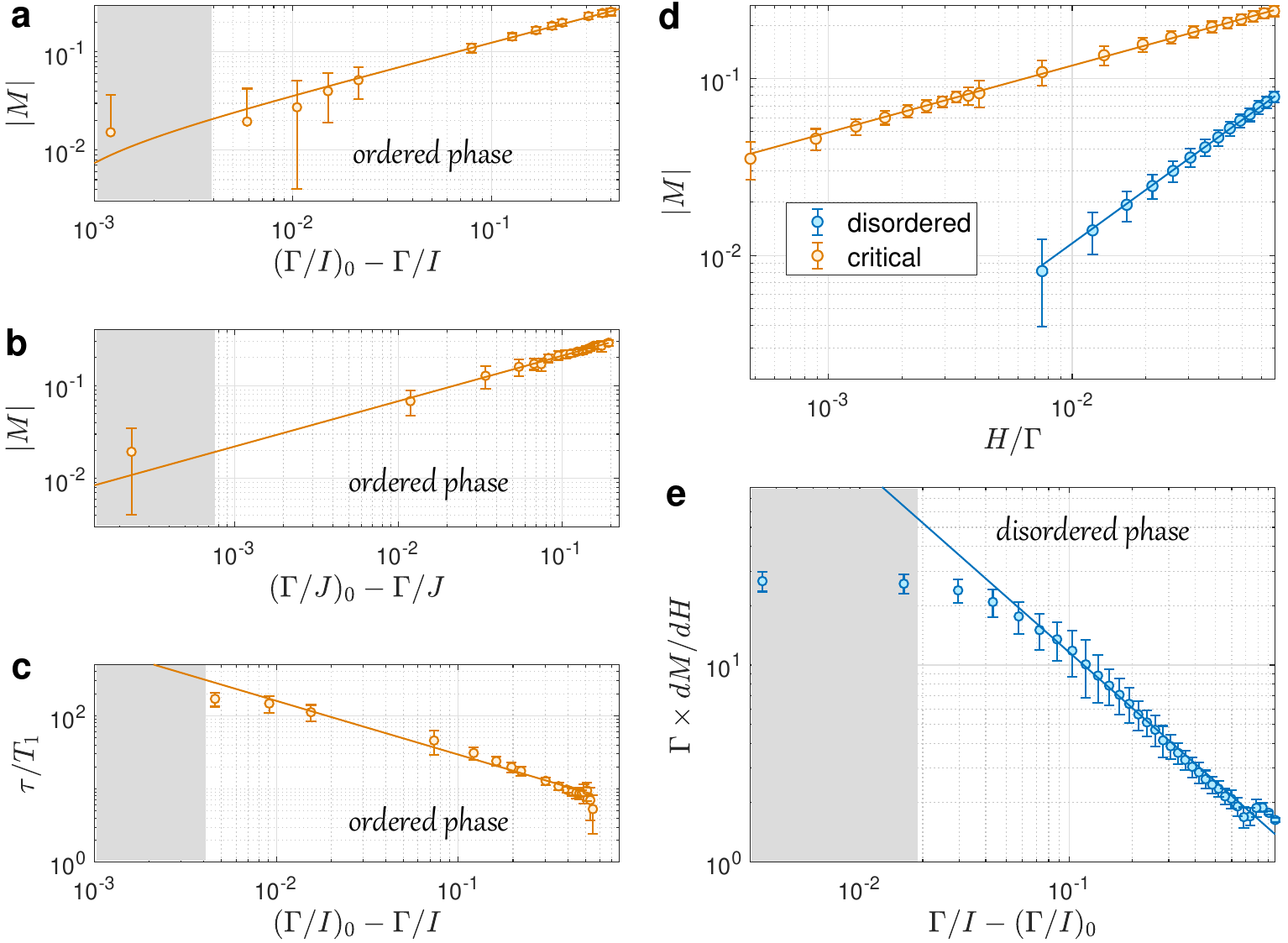}\setlength{\belowcaptionskip}{-4mm}\caption{\textbf{Critical exponents fitting in a logarithmic scale}. \textbf{a.} The critical exponent $\beta_I$. \textbf{b.} The critical exponent $\beta_J$. \textbf{c.} The critical exponent $z\nu$. \textbf{d.} The critical exponent $\delta$ obtained by fitting the data of the critical value $I=1.5\Gamma$ (orange), compared to standard optical-pumping with $I=0$ (blue). \textbf{e.} The critical exponent $\gamma$.
    \label{fig:averaged_I} }. 
\end{figure*}

\begin{figure*}[bt]    \centering
    \includegraphics[width=11.0 cm]{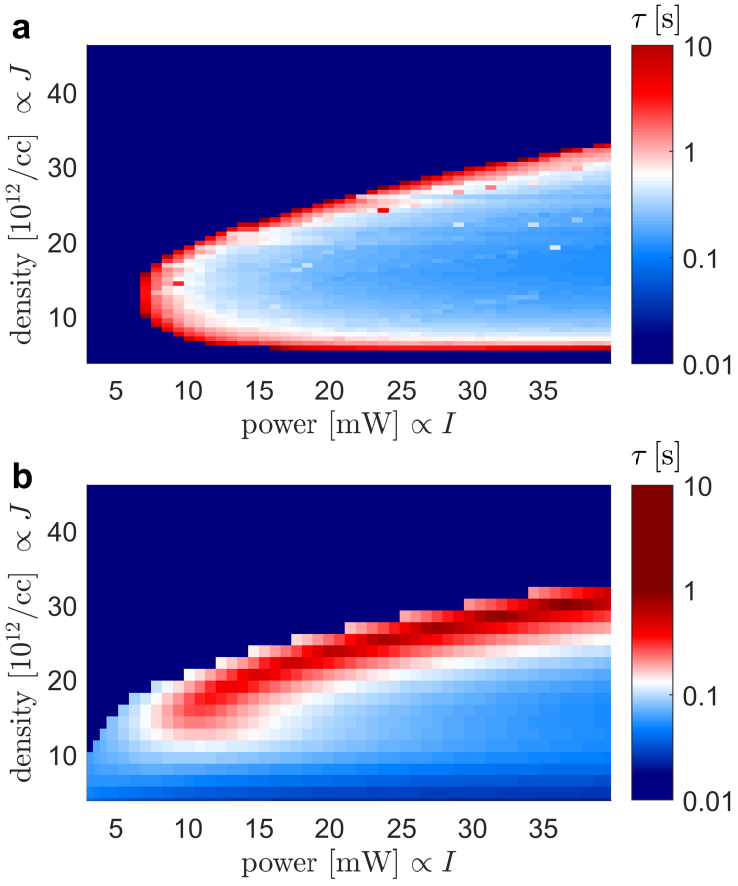}\setlength{\belowcaptionskip}{-4mm}\caption{\textbf{Dependence of the spin-response time on the Zeeman coherence}. \textbf{a.} Simulated spin-response time with zeroed Zeeman coherences. \textbf{b.} Simulated spin-response time at a small, finite magnetic field of $0.1\text{mG}$. For the experiment carried at $B=1$ G, the Zeeman coherences are relaxed by rapid spin-exchange collisions. \label{fig:magnetic_field} } 
\end{figure*}

\end{document}